\documentclass[fleqn,10pt]{wlscirep}
\usepackage{color}
\usepackage{xcolor}
\usepackage{epstopdf}
\title{Quantum synchronization and  quantum state sharing in irregular complex network}

\author[1]{Wenlin Li}
\author[1]{Chong Li}
\author[1,*]{Heshan Song}
\affil[1]{School of Physics and Optoelectronic Engineering, Dalian University of Technology, Dalian 116024, China}

\affil[*]{corresponding.hssong@aliyun.com}

\begin{abstract}
We investigate quantum synchronization phenomenon within the complex network constituted by coupled optomechanical systems and prove the unknown identical quantum states can be shared or distributed in the quantum network even though the topology is varying. Considering a channel constructed  by quantum correlation, we show that quantum synchronization can sustain and maintain high levels in Markovian dissipation for a long time. We analyze state sharing process between two typical complex networks, that is, a small-world network corresponding to linear motif state sharing and a scale-free network corresponding to whole network sharing, respectively. Our results predict that linked nodes can be directly synchronized in small-world network, but the whole network will be synchronized only if some specific synchronization conditions are satisfied. Furthermore, we give the synchronization conditions analytically through analyzing network dynamics. This proposal paves the way for studying multi-interaction synchronization and achieving an effective quantum information processing in complex network.
\end{abstract}
\begin{document}

\flushbottom
\maketitle
%
%
\thispagestyle{empty}
\section*{Introduction}
Synchronization is one of the most intriguing and valuable phenomena in classical physics and its history can be traced back to the observation of two pendulum clocks by Huygens in the 17th century \cite{1}. In the last decade, synchronization idea has been widely applied in the fields of control and communication \cite{2,3,4,4p}, which urges people to search for similar phenomena in quantum regime. Among them, a pioneering and significant progress is that Mari et al \cite{5}. extended the concept of complete synchronization into continuous variable (CV) quantum system and characterized it by a quantitative measure. Up to now, quantum synchronization has been paid extensive attention in many physical systems \cite{6,7,8,9,10,11,12,13}, but few works proposed it as a tool in view of applications. Recently, some effective attempts are presented to apply quantum synchronization in signal transmission \cite{14}, parameter identification \cite{15} and atomic clock \cite{7,16}. Owing to Heisenberg uncertainty \cite{1}, however, quantum effect appears to just take place a negative influence on synchronization behavior due to quantum fluctuation. The majority of previous works considered such kind of synchronization of only expectation value in quantum system and the quantum fluctuation is neglected or regarded as disturbance in their schemes \cite{7,14,15,16}. 

Intuitively, an appropriate application of quantum synchronization is to provide an effective quantum correlation for quantum information processing (QIP) \cite{17}. Different from the applications of synchronization in other fields, quantum characteristics play the important role in QIP. That is not just because synchronization means two systems take on homology evolutions, which indicates the information encryption and transmission between such two systems are convenient \cite{18,19}. Simultaneously, non-local quantum effect is indispensable in this process in order to obtain the particular security and efficiency of QIP. 

Other significant advantages of quantum synchronization are controllability and accessible extendibility. Especially in recent years, it is expected that QIP can be extended well into $n$-body scheme or a quantum network \cite{20,21,22,23}. However, the crossover between the quantum synchronization and complex network remains largely unexplored. In the past decade, quantum network protocols are based mainly on one-dimensional arrays or some regular networks in order to simplify or avoid the complex multi-interaction \cite{24,25,26}. Although the synchronization and correlation in a random network constituted by some simple physical systems (identical van der Pol oscillators, for example) have been discussed in few recent works \cite{28,29}, it still remains twofold difficult to establish a general quantum network by applying existing results. For one hand, processing quantum information needs more complex hybrid systems with higher dimensions and different (random) parameters (initial states). On the other hand, network theory has proved that some typical network structures (e.g. scale-free network and small-world network) are more accurate descriptions of actual information processing network compared to completely random structure \cite{30,31,32,33}.

The aim of our work is to address the above problems through proposing a QIP scheme based on the application of quantum synchronization and expanding QIP well in complex quantum network. Specifically in this paper, we study a quantum state sharing scheme (also called state distribution scheme) in the frame of optomechanical systems. It will be known that such a QIP process requires a genuine quantum synchronization channel since the shared quantum states need to couple with the channel directly. The synchronization channel is composed of oscillators which are twofold controlled by phonon and circuit couplings for eliminating the difference between the initial state and the dynamics parameter of each oscillator. This design allows us to obtain quantum synchronization between two completely different oscillators even in weak coupling range, however, the system accessing illegally into the network will not be synchronized with other systems because of the notable differences. 

Through further discussion about complex network theory, we determine that this quantum synchronization can also exist in a multi-node network in actual communication process, for instance, small-world (SW) or scale-free (SF). For these two kinds of networks, we will give the synchronization conditions analytically and ensure that the quantum synchronization and the state sharing will always effective even though the topology of network varies with time.

\section*{Quantum synchronization theory}
We consider two coupled quantum systems which can be completely described by the quadrature operators (e.g. dimensionless position operator $\hat{q}$ and momentum operator $\hat{p}$) in the Heisenberg picture. The difference between two systems can be characterized by the following defined error 
operators 
\begin{equation}
\begin{split}
\hat{q}_-(t)&\equiv [\hat{q}_1(t)-\hat{q}_2(t)]/\sqrt{2}\\
\hat{p}_-(t)&\equiv [\hat{p}_1(t)-\hat{p}_2(t)]/\sqrt{2},
\label{eq:clala}
\end{split}
\end{equation}
and quantum complete synchronization will be realized when $q_-$ and $p_-$ vanish asymptotically  with evolution. For further quantitative statement, we introduce a synchronization measure proposed by Mari et al., that is \cite{5,34}, 
\begin{equation}
S_c(t)={\langle{\hat{q}_-(t)}^{2}+{\hat{p}_-(t)}^{2}\rangle}^{-1}.
\label{eq:sc}
\end{equation}
Compared to previous works, $S_c$ is a good metric for genuine quantum synchronization because the influences of quantum fluctuation and nonlocal quantity are both considered in this synchronization measure, meaning that it can effectively distinguish the classical synchronization (even in quantum system) and genuine quantum synchronization.

For mesoscopic CV systems, $S_c$ can be modified as 
\begin{equation}
S'_c(t)={\langle{\delta\hat{q}_-(t)}^{2}+{\delta\hat{p}_-(t)}^{2}\rangle}^{-1}
\label{eq:scma}
\end{equation}
by mean-field approximation. Every operator here can be rewritten as a sum of its expectation value and a small fluctuation near the expectation value, i.e., it is redefined in the following form $\hat{o}_-=\langle\hat{o}_-\rangle+\delta \hat{o}_-$ with $o\in\{q,p\}$. Because we only ignore the expectation value of each operator in Eq. \eqref{eq:scma}, $S'_c$ contains all the quantum properties of $S_c$ and it can also be regarded as a quantum synchronization measure when $\lim_{t\rightarrow\infty}\langle \hat{o}_-\rangle=0$ which is exact classical synchronization condition is satisfied. Mathematically, this is because $S'_c$ will be equal to $S_c$ when $\lim_{t\rightarrow\infty}\langle \hat {o}_-\rangle=0$; And physically, synchronization in level of expected value can be regarded as a necessary condition of quantum synchronization. Therefore, $S'_c(t)$ is defined as second order quantum synchronization measure in following discussion to reflect quantum property differences between systems, correspondingly, $\langle \hat{o}_-\rangle$ is regarded as a first order measure to judge whether the expectation values are synchronous or not.

\section*{Dynamics of hybrid electro-optomechanical system}
\begin{figure}[]
\centering
\includegraphics[width=5.5in]{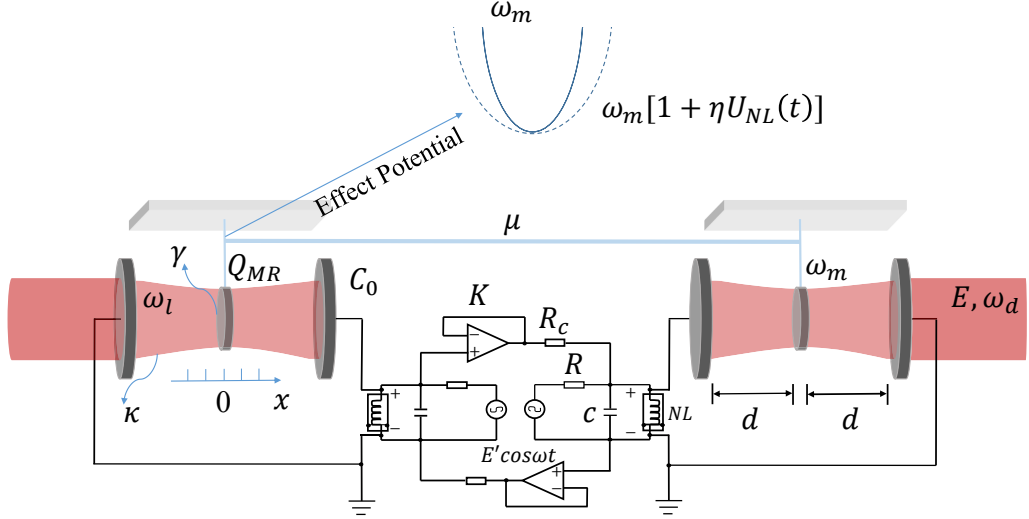}  
\caption{\textbf{Schematic diagram of point-to-point quantum state sharing}. Two hybrid electro-optomechanical systems are coupled via a phonon tunnelling and a linear resistor. For each subsystem, a charged oscillator is placed at wave node of a Fabry-P\'erot cavity and it couples with the cavity field via a linear optomechanical interaction. There exists an electric potential difference between walls of a cavity  which is provided by the inductance of the Duffing circuit. 
\label{fig:fig1}}
\end{figure}
Let us start by focusing on the dynamics of hybrid electro-optomechanical system. As schematically shown in Fig. \ref{fig:fig1}, the charged mechanical oscillators are coupled to the optical field and parametrically interact with the charged cavities which also play the role of electrodes. Two oscillators mutually couple through a phonon tunnelling and the electrode voltages are provided by two Duffing circuits coupled to each other via a linear resistor. We emphasize the electro-oscillator interaction is a parametric coupling because it can be thought as a deviation in respective potential terms of two oscillators. This effect can be regarded as a time-dependent rescaling of the mirror frequency \cite{35,36,37,38,39,40}. For a  freely moving oscillator corresponding to Hamiltonian $H_m=\hat{P}^2/2m+m\omega^2_m\hat{x}^2/2$, the modified Hamiltonian under the control of the bias gate becomes
\begin{equation}
H_m=\dfrac{\hat{P}^2}{2m}+\dfrac{1}{2}m\omega^2_{eff}\hat{x}^2, 
\label{eq:electro-oscillatorod}
\end{equation}
where $\hat{x}$ and $\hat{P}$ are the position and momentum operators of the oscillator with the bare eigen-frequency $\omega_m$ and the effective mass $m$. The effective frequency can be expressed as $\omega_{eff}=\omega_m(1+\eta U_{NL}(t))$, where $U_{NL}(t)$ is the voltage of nonlinear inductor and $\eta$ is a constant factor depending on parameters of circuit. By defining the non-dimensional coordinate and momentum operators $\hat{q}=\sqrt{m\omega_m}\hat{x}$ and $\hat{p}=\hat{P}/\sqrt{m\omega_m}$ and using relation $b=(\hat{q}+i\hat{p})/\sqrt{2}$, Eq. \eqref{eq:electro-oscillatorod} can be rewritten as 
\begin{equation}
H_m=\omega_mb^{\dagger}b+\dfrac{\omega_m}{4}\eta U_{NL}(t)(b^{\dagger}+b)^2\equiv\omega_mb^{\dagger}b+\dfrac{\omega_m}{4}C_j(t)(b^{\dagger}+b)^2.
\label{eq:electro-oscillator}
\end{equation}
where $b$ is phonon annihilation operator satisfying $[b,b^{\dagger}]=1$. Under the modified potential, the Hamiltonian corresponding to this model can be divided into three parts:
\begin{equation}
H=\sum_{j=1,2}H_{0j}+H_{int}+H_{ej},
\label{eq:add}
\end{equation}
where $H_{0j}=\omega_{lj}a_{j}^{\dagger}a_{j}+\omega_{mj}b^{\dagger}_{j}b_{j}-ga^{\dagger}_ja_j(b^{\dagger}_{j}+b_{j})+iE(a^{\dagger}_je^{-i\omega_{dj}t}-a_je^{i\omega_{dj}t})$ is the standard Hamiltonian of optomechanical system \cite{41,42}, $H_{int}=-\mu(b^{\dagger}_{1}b_{2}+b^{\dagger}_{2}b_{1})$ is phonon interaction through the tunnelling with intensity $\mu$ \cite{5,14}. $H_e$ describes the Coulomb interaction caused by two electrodes.

Here we provide the details behind Eq. \eqref{eq:electro-oscillator} by analyzing the dynamics 
of the electrical circuit system. A simple Duffing circuit can be described by following dynamics equation \cite{60}:
\begin{equation}
\begin{split}
\dfrac{d^2\phi}{d\tau^2}+\dfrac{1}{RC}\dfrac{d\phi}{d\tau}+\dfrac{{\chi}_1}{C}\phi+\dfrac{\chi_3}{C}\phi^3=\dfrac{E'}{RC}\cos\omega t.
\label{eq:Duffing}
\end{split}
\end{equation}
Here $\chi_1$ and $\chi_3$ are constants depending on the type of the inductor and they satisfy following relationship $i=i_R-i_C=\chi_1\phi+\chi_3\phi^3$. $\phi$ is the flux over inductor, moreover, $i_R$($i_L$) and $V_R$($V_L$) are the current and voltage of the resistor (inductor), respectively. Here we make dimensionless transformation by setting $\varphi=\phi/\phi_0$, $t=\tau\sqrt{\chi_1/c}$, $U_{NL}=d\varphi/dt$, $\upsilon=\phi_0^2\chi_3/\chi_1$, $\varepsilon=(R\sqrt{\chi_1C})^{-1}$, $\mathbb{E}=E'/\chi_1R\phi_0$ and $\omega_0=\omega\sqrt{C/\chi_1}$.
In this picture, the unidirectional coupling via a linear resistor can be described as a control term $\varepsilon K(U^{con}_{NL}-U^{self}_{NL})$, where $K$ is coupling intensity. Therefore, for $j=1,2$, two mutual controlled Duffing circuits in Fig. \ref{fig:fig1} can be expressed as:
\begin{equation}
\begin{split}
&\dfrac{d}{dt}\varphi_j=U_{NL,j}\\
&\dfrac{d}{dt}U_{NL,j}=-\varepsilon U_{NL,j}-\varphi_j-\upsilon\varphi_j^{3}+\mathbb{E}\cos\omega_0 t+\varepsilon K(U_{NL,3-j}-U_{NL,j}).
\label{eq:Duffingcoupling}
\end{split}
\end{equation}

The Coulomb interaction provides an additional potential energy ${\omega_m}\eta' V_L(b^{\dagger}+b)^2/4$ to the oscillator, which has been deduced in Ref. \citen{38}. Through utilizing similar dimensionless transformation, the additional potential energy corresponding to the $j$th oscillator can be gained
\begin{equation}
H_{ej}=\dfrac{\omega_mj}{4}\eta U_{NL,j}(b^{\dagger}+b)^2,
\label{eq:heff}
\end{equation}
with the characteristic parameter
\begin{equation}
\eta=\sqrt{\dfrac{\chi_1}{c}}\phi_0\eta'=\dfrac{C_0Q_{MR}\phi_0}{\pi\varepsilon_0m\omega_m^2d^3}\sqrt{\dfrac{\chi_1}{C}}.
\label{eq:contorlexex}
\end{equation} 
Then the total Hamiltonian of this system can be expressed as ($\hbar=1$)
\begin{equation}
H=\sum_{j=1,2}\left\{-\Delta_ja_{j}^{\dagger}a_{j}+\omega_{mj}[1+\dfrac{C_j(t)}{2}]b^{\dagger}_{j}b_{j}-iga^{\dagger}_ja_j(b^{\dagger}_{j}+b_{j})+iE(a_{j}^{\dagger}-a_{j})+\dfrac{\omega_{mj}}{4}C_j(t)(b^{\dagger}_{j}b^{\dagger}_{j}+b_{j}b_{j})\right\}-\mu(b^{\dagger}_{1}b_{2}+b^{\dagger}_{2}b_{1}),
\label{eq:Hamtion}
\end{equation}
after a frame rotating. Here for $j=1,2$, $a_j$ ($a^{\dagger}_j$) and $b_j$ ($b^{\dagger}_j$) are the optical and mechanical annihilation (creation) operators. $\Delta_j=\omega_{dj}-\omega_{lj}$ refers to the detuning between the frequencies belonging respectively to the laser drive and the cavity mode. $\omega_{mj}$ is the mechanical frequency. $g$ is the optomechanical coupling constant and $E$ is the drive intensity.    

Based on the Hamiltonian in Eq. (\ref{eq:Hamtion}), we consider the dissipative effects in the Heisenberg picture and write the quantum Langevin equations as follows \cite{43,35}:
\begin{equation}
\begin{split}
&\dot{a}_j=[-\kappa+i\Delta_j+ig(b^{\dagger}_j+b_j)]a_j+E+\sqrt{2\kappa}{a}^{in}_{j}\\ 
&\dot{b}_j=\left\{-\gamma-i\omega_{mj}[1+\dfrac{C_j(t)}{2}]\right\}b_j+iga^{\dagger}_{j}a_{j}+i\mu b_{3-j}-i\dfrac{\omega_{mj}}{2}C_j(t)b^{\dagger}_j+\sqrt{2\gamma}{b}^{in}_{j}.
\label{eq:qle}
\end{split}
\end{equation}
In this expression, $\kappa$ and $\gamma$ are the optical and mechanical damping rates respectively, ${a}^{in}_{j}$ and ${b}^{in}_{j}$ are the input bath operators. Under the Markovian approximation, the input  operators are assumed to be white Gaussian fields obeying standard correlation, that is, $\langle{a}^{in,{\dagger}}_{j}{(t)}{a}^{in}_{j'}{(t')}+{a}^{in}_{j'}{(t'){a}^{in,{\dagger}}_{j}{(t)}}\rangle=\delta_{jj'}\delta(t-t')$ \cite{45} and $\langle{b}^{in,{\dagger}}_{j}{(t)}{b}^{in}_{j'}{(t')}+{b}^{in}_{j'}{(t'){b}^{in,{\dagger}}_{j}{(t)}}\rangle=(2\bar{n}_b+1)\delta_{jj'}\delta(t-t')$, where $\bar{n}_b=[\exp ({\hbar\omega_mj}/{k_BT})-1]^{-1}$ is the mean phonon number of the mechanical bath, which gauges the temperature $T$ \cite{46}. 

Here we adopt mean--field approximation to simplify above nonlinear differential operator equations since it is quite difficult to directly solve them \cite{5,13,47,48,49}. Therefore, each operator in quantum Langevin equation is expanded as the sum of a $c$ number mean value and a fluctuation operator, that is, $a_j(t)=\langle a_j(t)\rangle+[a_j(t)-\langle a_j(t)\rangle]:=A_j(t)+\delta a_j$ and $b_j(t):=B_j(t)+\delta b_j$. Under a strong
laser drive, the fluctuation can be regarded as perturbation around the corresponding mean value. In this case, Eq. (\ref{eq:qle}) can be divided into two different sets of equations , that is, for the mean value:
\begin{equation}
\begin{split}
&\dot{A}_j=[-\kappa+i\Delta_j+ig(B^{*}_j+B_j)]A_j+E\\ 
&\dot{B}_j=\left\{-\gamma-i\omega_{mj}[1+\dfrac{C_j(t)}{2}]\right\}B_j+ig\vert A_{j}\vert^2+i\mu B_{3-j}-i\dfrac{\omega_m}{2}C_j(t)B^{*}_j,
\label{eq:meanvalue}
\end{split}
\end{equation}
and for the fluctuation: 
\begin{equation}
\begin{split}
&\delta\dot{a}_j=[-\kappa+i\Delta_j+ig(B^{*}_j+B_j)]\delta a_j+igA_j(\delta b^{\dagger}_j+\delta b_j)+\sqrt{2\kappa}{a}^{in}_{j}\\ 
&\delta\dot{b}_j=\left\{-\gamma-i\omega_{mj}[1+\dfrac{C_j(t)}{2}]\right\}\delta b_j+igA_j\delta a^{\dagger}_j+igA^{*}_j\delta a_j +i\mu \delta b_{3-j}-i\dfrac{\omega_m}{2}C_j(t)\delta b^{\dagger}_j+\sqrt{2\gamma}{b}^{in}_{j}.
\label{eq:fluctuations}
\end{split}
\end{equation}
In above expressions, the quantum fluctuations have been already linearized by neglecting all second order terms. Utilizing mean-field approximation, except convenience, is also based on following two considerations. One is that quantum synchronization measure is of a clearer physical significance. Here we wish to emphasize once again that the synchronization in level of expected value in this case can be regarded as a necessary condition of quantum synchronization. The other is that mean--field approximation neglects the nonlinear effect in quantum level, which causes the quantum properties of the system are restricted to linear transformation. In our work, this characteristic can ensure that the system is always a Gaussian state \cite{50}. It is the reason why the Gaussian fidelity and the Gaussian entanglement are accurate in following discussion.

\section*{Point-to-point quantum synchronization and state sharing}
We firstly consider point-to-point quantum state sharing between two systems connected directly  
(see Fig. \ref{fig:fig1}). Two cavities here are the carriers of quantum states and we hope to prepare identical unknown states via quantum synchronization oscillators. Because of the linearization for the quantum fluctuation, the systems can be described by Eqs. \eqref{eq:Duffingcoupling}, \eqref{eq:meanvalue} and \eqref{eq:fluctuations} completely. By computing the covariance matrix of the systems (See \textbf{Appendix} for details), the evolution of synchronization measure in Eq. \eqref{eq:scma} can be calculated conveniently. Because the oscillators are directly coupled, we use non--local measure $S'_c$ to describe oscillator synchronization. For cavity fields, quantum sharing is more concerned with the consistency of the local quantum states in each cavity. Hence, we use local measure fidelity to describe optical fields.

\begin{figure}[]
\centering
\includegraphics[width=6in]{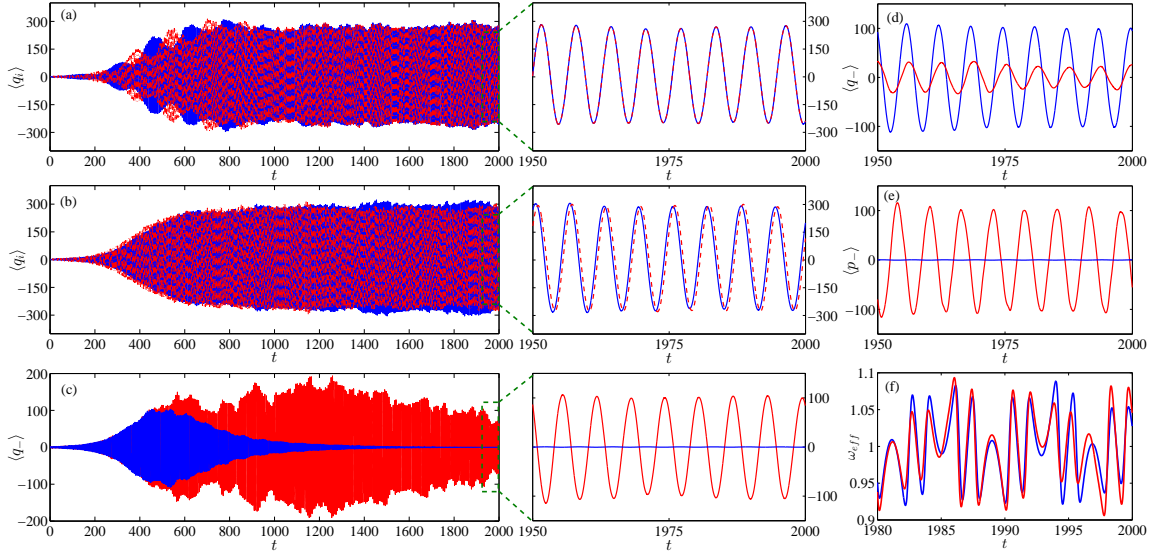}  
\caption{\textbf{Evolutions of expectation values.} (a, b): Expectation values of the oscillator coordinates corresponding to existing or disconnecting of coupling, respectively. Here blue solid lines denote $\langle q_1\rangle$ and red dot lines denote $\langle q_2\rangle$. (c, d): Expectation values of the  coordinate error operator respectively corresponding to $K=2$, $\mu=0.02$ (c, blue); $K=0$, $\mu=0$ (c, red); $K=2$, $\mu=0$ (d, blue) and $K=0$, $\mu=0.02$ (d, red). (e): Expectation values of the  momentum error operator corresponding respectively to $K=2$, $\mu=0.02$ (blue) and $K=0$, $\mu=0$ (red). (f): Effective frequencies of system $1$ (bule) and $2$ (red) when $K=0$. In these simulations, the oscillator frequency is set $\omega_{m1}=1$ as a unit and other parameters are: $\omega_{m2}=1$, $\omega_0=0.8$, $\Delta_j=\omega_{mj}$, $g=0.005$, $\kappa=0.15$, $\gamma=0.005$, $\eta=0.01$, $\varepsilon=0.18$, $\nu=1$, $E=10$ and $\mathbb{E}=26.7$. The initial state of the cavity field is vacuum state which corresponds to $A_j(0)=0$ and other initial conditions are all random. 
\label{fig:fig2}}
\end{figure}
In Fig. \ref{fig:fig2}, we show dynamics of oscillators by plotting the evolutions of the operator expectation values. One can obviously see that synchronous evolution between the oscillators  appears under the suitable coupling intensity (Fig. \ref{fig:fig2}(a)), but this synchronization will be destroyed when both two couplings are disconnected (Fig. \ref{fig:fig2}(b)) and two evolution curves are inconsistent. In Fig. \ref{fig:fig2} (c) and (d), we explain this phenomenon more intuitively by considering the first-order error $\langle q_-\rangle=\langle q_1\rangle-\langle q_2\rangle$. Here $\langle q_-\rangle$ is plotted under different connections and it shows that the error will always tend to zero if and only if two couplings exist simultaneously (blue line in (c)). Otherwise, the error will take on irregular evolution with large amplitude ( (d) and red line in (c) ). Fig. \ref{fig:fig2}(e) illustrates that all the conclusions obtained from the generalized coordinate can also be applied to the generalized momentum. In Fig. \ref{fig:fig2}(f), we plot the effective frequency $\omega_{eff}$ of each subsystem to illustrate the  significant difference between two systems when $K=0$.   

\begin{figure}[]
\centering
\includegraphics[width=6in]{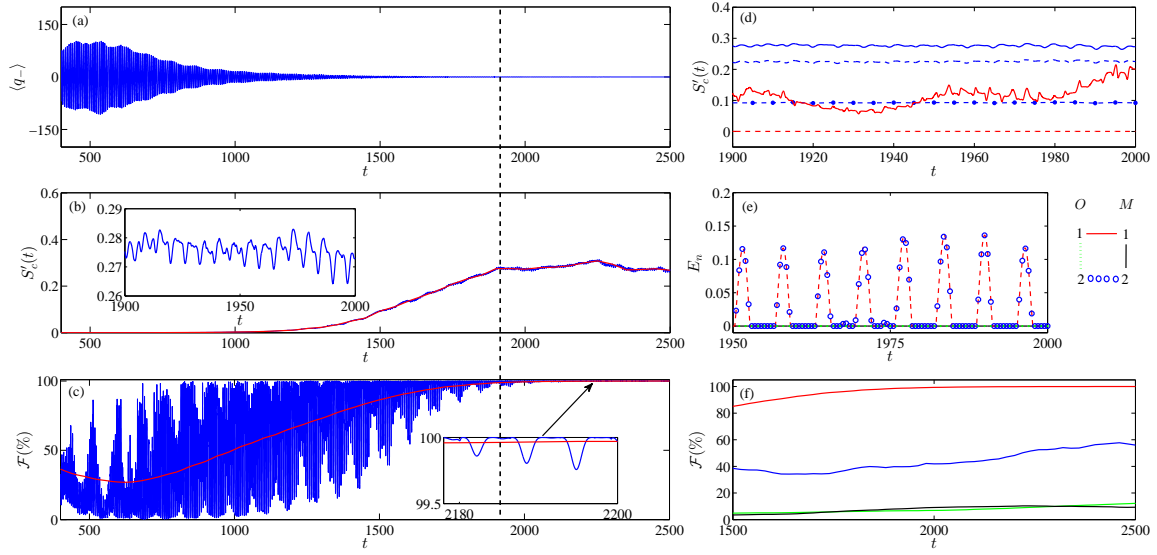}  
\caption{\textbf{Analyses of the quantum synchronization evolutions.} (a, b, c): Comparison among the error of expectation value $\langle q_-\rangle$, quantum synchronization measure $S'_c(t)$ and Gaussian fidelity $\mathcal{F}$. The inset in (b) is the partial enlarged drawing of $S'_c$ in $t\in[1900,2000]$. In (b) and (c), the red lines express the local average synchronization measure and Gaussian fidelity by calculating  $\xi(\bar{t})={\Delta t}^{-1}\int_t^{t+\Delta t}\xi dt$ in the time window $\Delta t$ ($\xi\in\{S'_c,\mathcal{F}\}$);  (d): Quantum synchronization measures under different conditions. Here blue lines denote $S'_c$ when $\bar{n}_b=0$ (solid), $\bar{n}_b=0.25$ (dotted) and $\bar{n}_b=2.5$ (circle). Red lines denote $S'_c$ under $K=0$, $\mu=0.02$ (solid) and $K=0$, $\mu=0$ (dotted). (e): Evolution of \textit{logarithmic negativity}. (f): Local averaged fidelities respectively corresponding to $K=2$, $\mu=0.02$ (red); $K=2$, $\mu=0$ (green); $K=0$, $\mu=0.02$ (blue) and $K=0$, $\mu=0$ (black). Here all the other parameters are the same with Fig. \ref{fig:fig2}.
\label{fig:fig3}}
\end{figure}
While $\langle q_-\rangle\rightarrow 0$ and $\langle p_-\rangle\rightarrow 0$ are simultaneously satisfied, two oscillators will exhibit the characteristics of complete synchronization in the level of expectation value. In some previous works, this kind of synchronization is also considered as a quantum synchronization. This view is not strict because expectation value is a incomplete description without considering quantum fluctuation. If the synchronization is not intended to be used in semi-classical information processing (e.g., transmission of strong signal and parameter identification in quantum ensemble) but needs more quantum properties, the nonlocality and quantum fluctuation will also cause a critical impact and they can not be ignored in quantum synchronization analyses. In particular for Gaussian state sharing discussed in our work, a genuine quantum synchronization will be required. 

In Fig. \ref{fig:fig3}, we present the contrast among the expectation value synchronization, quantum synchronization and Gaussian fidelity \cite{51,52}. Compared with only considering the expectation value, here evolution of the system is more complicated and it can be subdivided to three distinguishable processes according to the quantum synchronization measure $S'_c$. Firstly, two systems are not synchronized because there obviously exists classical error between oscillator expectation values. Correspondingly, $S'_c$ always tends to $0$. Subsequently,  with the classic error gradually tending to zero, $S'_c$ also gradually increases with time evolution at this stage. 
The gradually rising $S'_c$ illustrates that this process is a transformation process from quantum non-synchronization to quantum synchronization. Physically, the appearance of this process is due to the mutual modulation between two systems. From Fig. \ref{fig:fig3}(c), we find that the corresponding fidelity of two cavity fields is also rising in this duration, but it is not yet available for quantum sharing. With the evolution continuing and finally for  $S'_c$, an inflection point appears  (see dotted boundary line). After this inflection point, stable non-zero $S'_c$ emerges and maintains in a long time interval and the corresponding fidelity tends to $100\%$, which implies two cavity fields evolve gradually from vacuum states to the quantum states with almost $100\%$ reliability. The inset in Fig. \ref{fig:fig3}(b) shows that $S'_c$ is greater than $0.26$ after the inflection point, which is a higher value compared to Mari's results. It can be sure from above analyses that  quantum fluctuation can be regarded as synchronization in this process.

In Fig. \ref{fig:fig3}(d), we show quantum synchronization measures under different conditions. The results show that $S'_c$ will decrease to $0.1$ when the classical coupling is disconnected. Although $S'_c$ is unequal to zero in this case, it does not mean that the systems have been synchronized because the classical error will no longer tend to zero (see Fig. \ref{fig:fig2}(d)). Moreover, $S'_c$ will equal to zero if both couplings are disconnected. By comparing these two results, it can be proved that the quantum coupling is more suitable for playing a role in restraining difference between the quantum fluctuations. Similar conclusions can also be verified by Fig .\ref{fig:fig3}(f) which shows that $\mathcal{F}(\infty)\rightarrow 100\%$ will be satisfied only while both two couplings are connected synchronously. Fig. \ref{fig:fig3}(d) also shows that this synchronization will keep high efficiency if the bath temperature is limited to be lower than $T=1$mK (corresponding to MHz phonon frequency). Moreover, $S'_c$ will still be greater than $0.1$ even $T=5$mK, which corresponds to a strong robustness. Another concern in this work is whether the quantum properties of two systems are also identical at the synchronization moment. Therefore, we plot Fig. \ref{fig:fig3}(e) and confirm that the entanglement of two systems also takes on consistent evolution. Considering above properties together, we can finally determine that the synchronization between two systems indeed belongs to a genuine quantum synchronization. Fig. \ref{fig:fig3}(e) also illustrates the two systems are always separable in both optical field and oscillator freedoms. This characteristic is suitable for quantum network because other nodes will be not disturbed by entanglement steering when a node is attacked or bugged.

Fig. \ref{fig:fig2}(c), (d) and Fig. \ref{fig:fig3}(f) illustrate that the couplings via phonon channel and linear resistor in our model are both necessary for synchronization. This feature is rarely used to design synchronization schemes in previous works. In fact, these two kinds of couplings respectively exhibit different physical mechanisms in the synchronization process. Here quantum coupling plays a similar role with that reported in Ref. \citen{13}, i.e.,  it is used to eliminate the initial difference between two identical systems by mutual adjustment, and to generate nonlocality in order to improve synchronization measure. However, different from Ref. \citen{13}, every subsystems in our work can be treated as the same if and only if circuit coupling is connected. On the other words, the synchronization mechanisms in our paper are as follows: Circuit coupling controls subsystems to eventually have the same effective frequency, which means two subsystems evolve under the same dynamics equation. Once this condition is satisfied, the responsibility of phonon coupling is to offset the initial difference between two subsystems. On the contrary, if circuit coupling is disconnected or one of the subsystems is a common optomechanical system without control voltage ($\omega_{eff}=\omega_m$), dynamics equations of the two systems are different and exhibit unequal effective frequencies (see Fig. \eqref{fig:fig2}(f)). In this case, weak quantum coupling will not be sufficient to synchronize two different systems. Compared to related works which obtain quantum synchronization by only one coupling, our scheme can ensure a wide range of achievable parameters since the quantum coupling is only responsible for synchronizing the initial difference. At the same time, two kinds of couplings can provide an additional control mode. Both the wider range of parameters and the additional control mode can make constructing of network more convenient.

\begin{figure}[]
\centering
\includegraphics[width=6in]{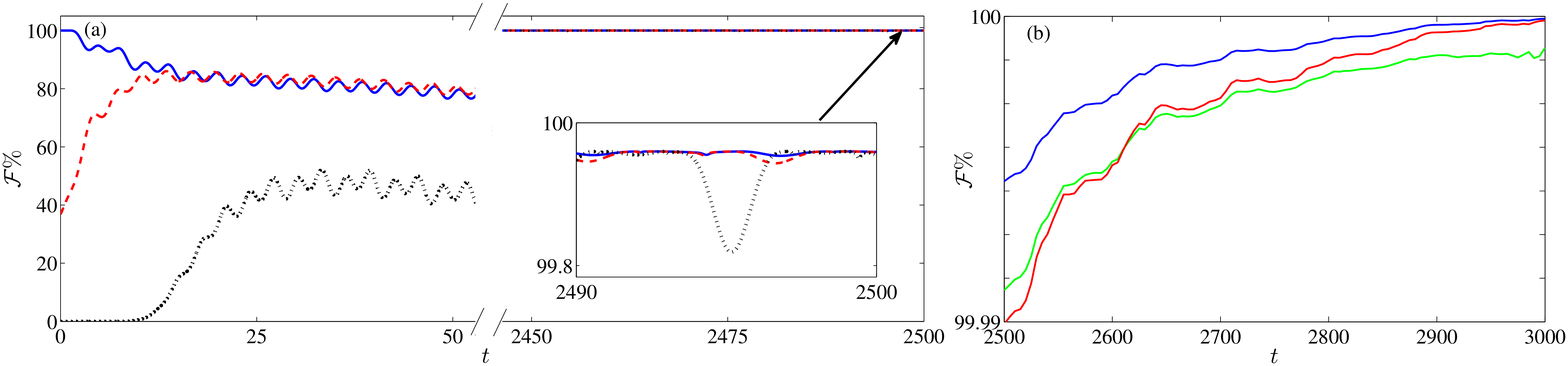}  
\caption{\textbf{Discussions of fidelity.} (a): Initial state dependency of fidelity. Here the initial states of cavity fields are respectively selected as two vacuum states: $\vert 0\rangle$ and $\vert 0\rangle$  (blue);  coherent states with identical photon number:  $\vert \alpha\rangle$ and  $\vert \alpha e^{i\phi}\rangle$ (red) and  coherent states with different photon number:  $\vert \alpha_1\rangle$ and  $\vert \alpha_2\rangle$ (black). Here $\vert\alpha\rangle=\exp(\alpha a^\dagger-\alpha^{*}a)\vert 0\rangle$, $\alpha=\alpha_1=1$, $\alpha_2=10$ and $\phi=\pi/2$. (b): Gaussian fidelities when $\Delta\bar{n}_b=\bar{n}_{b1}-\bar{n}_{b2}=0$ (blue), $\Delta\bar{n}_b=0.25$ (green) and $\Delta\bar{n}_b=0.5$ (red). All the other parameters are the same with Fig. \ref{fig:fig2}.
\label{fig:fig4}}
\end{figure}
To show the efficiency of the state sharing, finally, we end the analyses of point to point system with discussing the initial state dependence and environment influence  on Gaussian fidelity. In Fig. \ref{fig:fig4}(a), we plot the evolutions of the fidelity with different initial states. It can be known that $\mathcal{F}(\infty)\rightarrow 100\%$ can always be achieved even starting evolution from arbitrary initial state. This is because the synchronization effect is an intrinsic property of the system and it is irrelevant with the initial selection. Fig. \ref{fig:fig4}(b) illustrates that the cavity field fidelity will always tend to a hundred percent even though the corresponding oscillators are dissipating into different baths. This property is quite different with $S'_c$ which takes on an obvious decline once the phonon number within bath increases. This is due to the fact that the influence of bath on the system has be simplified as a decay parameter under the Markovian approximation and such kind of parametric difference is also balanced by the quantum coupling when the optomechanical interaction and system-bath interaction are both weak. This performance will relax the requirement for the experimental conditions, in other words, it is more feasible to extend idea of quantum state sharing from  the point to point system to the quantum network. 

\section*{Quantum synchronization and state sharing in complex network}
Now let us extend above conclusions about point-to-point system to analyze state sharing within the network. Similarly to the discussion about point to point system, we also begin this section with a dynamics analysis of hybrid electro-optomechanical system array. According to Eq. \ref{eq:Hamtion}, whole Hamiltonian of a quantum network can be expressed as:
\begin{equation}
\begin{split}
H=&H_{free}+H_{couple}\\
=&\sum^N_{i=1}\left\{-\Delta_ia_{i}^{\dagger}a_{i}+\omega_{mi}[1+\dfrac{C_i(t)}{2}]b^{\dagger}_{i}b_{i}-iga^{\dagger}_ia_i(b^{\dagger}_{i}+b_{i})+iE(a_{i}^{\dagger}-a_{i})+\dfrac{\omega_{mi}}{4}C_j(t)(b^{\dagger}_{i}b^{\dagger}_{i}+b_{i}b_{i})\right\}-\sum_{t,q}\mu_{jk}(b^{\dagger}_{j}b_{k}+b^{\dagger}_{k}b_{j}),
\label{eq:networkHamtion}
\end{split}
\end{equation}
where $N$ is the total number of nodes and $\mu_{jk}$ represents the coupling
strength between nodes $j$ and $k$. Here $C_i(t)$ can be determined by Duffing circuit equations
\begin{equation}
\begin{split}
&\dot{\varphi}_i=U_{NL,i}\\
&\dot{U}_{NL,i}=-\varepsilon U_{NL,i}-\varphi_i-\upsilon\varphi_i^{3}+\mathbb{E}\cos\omega_0 t+\sum^{k=i}_{t,c} \varepsilon K_{jk}(U_{NL,j}-U_{NL,i}),
\label{eq:Dadnetwork}
\end{split}
\end{equation}
correspondingly, other mechanical quantities satisfy following quantum Langevin equations: 
\begin{equation}
\begin{split}
&\dot{a}_j=[-\kappa+i\Delta_j+ig(b^{\dagger}_j+b_j)]a_j+E+\sqrt{2\kappa}{a}^{in}_{j}\\ 
&\dot{b}_j=\left\{-\gamma-i\omega_{mj}[1+\dfrac{C_j(t)}{2}]\right\}b_j+iga^{\dagger}_{j}a_{j}-i\dfrac{\omega_{mj}}{2}C_j(t)b^{\dagger}_j+\sum^{j=i}_{t,q}i\mu_{jk} b_{k}+\sqrt{2\gamma}{b}^{in}_{j}.
\label{eq:neadqle}
\end{split}
\end{equation}
After mean-field approximation, Eqs. \eqref{eq:neadqle} becomes
\begin{equation}
\begin{split}
&\dot{A}_i=[-\kappa+i\Delta_i+ig(B^{*}_i+B_i)]A_i+E\\ 
&\dot{B}_i=\left\{-\gamma-i\omega_{mi}[1+\dfrac{C_i(t)}{2}]\right\}B_i+ig\vert A_{i}\vert^2-i\dfrac{\omega_m}{2}C_i(t)B^{*}_i+\sum^{j=i}_{t,q}i\mu_{jk} B_{k},
\label{eq:Daanetwork}
\end{split}
\end{equation}
to describe the evolutions of expected values. Here we define two graph matrices, i.e., $G^c_{jk}=K_{jk}$ and $G^q_{jk}=\mu_{jk}$ represent the coupling structures of circuit coupling and phonon coupling, respectively. $G^c$ can be arbitrary matrix since circuit coupling is classical, however, the Hermitian Hamiltonian requires $\left(G^q\right)^\top=G^q$. Based on this expression, Eqs. \eqref{eq:Dadnetwork} and \eqref{eq:neadqle} can be rewritten in more compact form by using graph matrices \cite{55}:
\begin{equation}
\mathbf{\dot{X}}=F(\mathbf{X})+G^c(t)\otimes H_c\cdot\mathbf{X}+iG^q(t)\otimes H_q\cdot\mathbf{X}.
\label{eq:adnnetdynamics}
\end{equation}
In this expression, $\mathbf{X}=(\mathbf{X}^1, \mathbf{X}^2, \mathbf{X}^3..., \mathbf{X}^N)^{\top}$ is defined as a network tensor, where $\mathbf{X}^j=(\phi_j,U_{NL,j},A_j,B_j)$. The second term and the third term in the right side respectively correspond to classical coupling term $\sum_{t,c} \varepsilon K(U_{NL,j}-U_{NL,i})$ and quantum coupling $\sum^{j=i}_{t,q}i\mu_{jk} B_{k}$, and $F(\textbf{X})$ describing free evolution of each node includes the rest parts of Eqs. \eqref{eq:Dadnetwork} and \eqref{eq:neadqle}. $H_c$ and $H_q$ refer to the autocorrelation function which describe the nexus between the variables in the same node. If variable order is defined as $\mathbf{X}^j=(\phi_j,U_{NL,j},A_j,B_j)$, the  autocorrelation functions should be $H_c=\text{diag}(0,1,0,0)$ and $H_q=\text{diag}(0,0,0,1)$. 

With the advances of network theory, it is gradually known that the irregular but incompletely random network is better and more practical than other network structures for the application in communication or calculation process.
 In the last decades, SW network and SF network are widely investigated and can act as effective simulations of actual communication networks. This motivates us to propose synchronization schemes for SW network and SF network in following subsections, respectively. The SW network corresponds to the case that only few directly connected nodes in network are achieved for synchronization. Besides, it is not required to design additional control for synchronization in SW network. If the synchronization target is used to synchronize all nodes in irregular network, we need a designed synchronization condition to adjust the parameters of each node, this is the goal of the synchronization in SF network.

\subsection*{Synchronization in small-world network}
\begin{figure}[]
\centering
\includegraphics[width=6in]{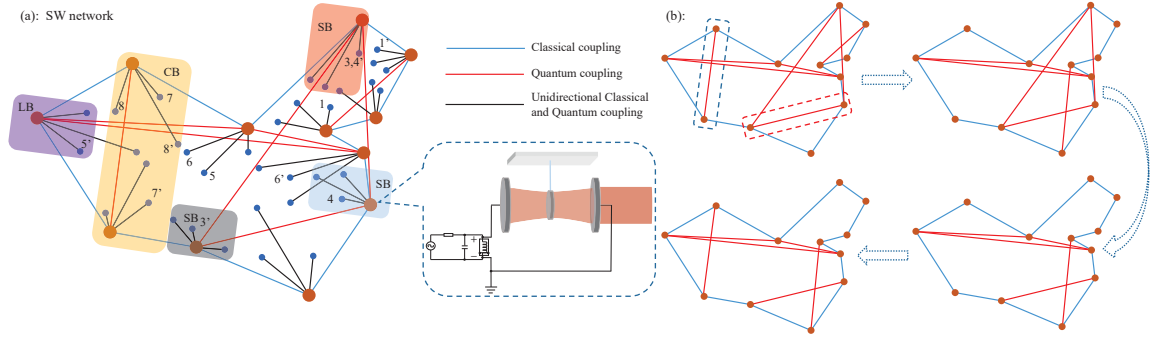}  
\caption{\textbf{Schematic diagram of SW network}. Each node represents a eletro-optomechanical system. The nodes connect to each other through different coupling forms and dissipate into different baths, including separate bath (SB), common bath (CB) and local bath (LB). Here (a) represents the structure of the SW network consisted of primary and secondary nodes and (b) shows a SW network with varying structure.
\label{fig:fig5}}
\end{figure}
We now go into detail about quantum synchronization in SW network. A typical Newman-Watts SW network can be regarded as to add few irregular links in the frame of a regular network with neighboring links. The construction of such  a network can be divided into two steps: 
\begin{itemize}
\item \textbf{Establishing regular network}: Staring from a ring-like network with regular connectivity comprising $N$ nodes and each node within the network connects to its $M$ (even number) nearest neighbors.
\item \textbf{Randomization edge adding}: In addition to the above links, each node can also connect to non-nearest neighbor nodes at random with small probability $P$.
\end{itemize}
For hybrid electro-optomechanical systems, above two structures can respectively correspond to the classical coupling and quantum coupling. As shown in Fig. \ref{fig:fig5}(a), we construct a SW network with following strategy: $N$ electro-optomechanical systems (orange points) are selected as the nodes of the network, and their circuit parts link to the nearest-neighbor nodes via classical Duffing couplings (blue lines); A node will not link any other nodes via quantum couplings (red lines) unless it wants to synchronize with other nodes. In our small-world network, time-dependent topology refers to that a node can select different nodes for quantum coupling in the time evolution. (see Fig. \ref{fig:fig5}(b)) 

\begin{figure}[]
\centering
\includegraphics[width=6in]{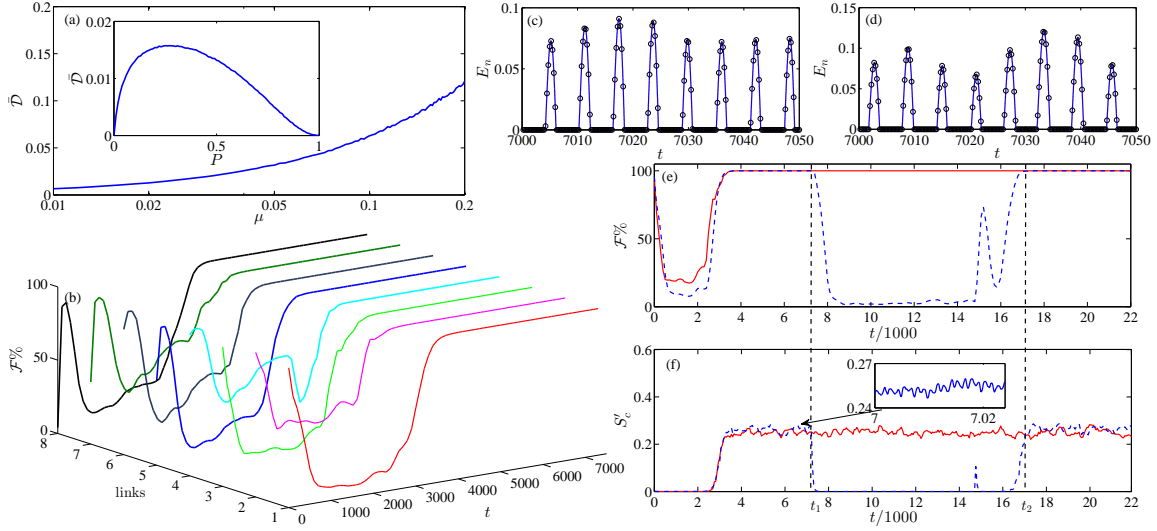}  
\caption{\textbf{Evolutions of fidelity and entanglement in SW network.} (a): Average trace 
distance $\bar{D}$ with varied coupling intensity $\mu$ under $N=12$, $M=2$ and $P=0.1$. The inset in (a) shows change of $\bar{D}$ with varied edge adding probability $P$ under $N=12$, $M=2$ and $\mu=0.02$. (b): Fidelities between two nodes in different linear motifs. Here link $i$ refers to the connection between nodes $i$ and $i'$ in Fig. \ref{fig:fig5}(a). (c) and (d): Entanglement measures of the quantum states in different cavities. (e) and (f): Fidelities and synchronization measures between two nodes when the network structure varies. What the  blue line represents is the link disconnects at 
$t_1$ and reconnects at $t_2$. The initial state of each cavity field is vacuum state or random coherent state. Other initial conditions are all random and all the other parameters are the same with Fig. \ref{fig:fig2}.
\label{fig:fig6}}
\end{figure}
In the case of larger $N$, the quantum coupling here can be regarded as random connection with a small probability $P$. Previous works have  proved that such network topology is of smaller average path length and larger clustering coefficient \cite{53,54}. To some extent, SW network remains symmetry like regular network and the node differences caused by network topology are small enough to be eliminated becasue of the weak coupling in quantum domain. Especially when $N$ and $P$ are fixed, the change of network structure will have tiny impact on direct connected nodes, which ensures that quantum synchronization can be extended more conveniently into SW network. 

Now we discuss above analyses in mathematics. Let us re-examine the dynamics equation in Eq. \eqref{eq:adnnetdynamics}. In view of  SW network, elements of $G^c$ should be zero except $g^c_{i,i\pm 1}=\varepsilon K$ and $g^c_{1,N}=g^c_{N,1}=\varepsilon K$. Correspondingly, elements of $G^q$ should perform as $g^q_{jk}=\mu_{jk}$. Substituting $G^c$ and $G^q$ into Eq. \eqref{eq:adnnetdynamics}, the dynamics of whole network can be determined by coupling matrix 
\begin{equation}
G_{N,M,P}(t)=G^c(t)\otimes H_c+iG^q(t)\otimes H_q.
\label{eq:ouhejuzhen}
\end{equation}
We can calculate the trace distance under different time
\begin{equation}
D(t,t')=\dfrac{1}{2}\text{Tr}\vert\dfrac{G_{N,M,P}(t)}{\vert G_{N,M,P}(t)\vert}-\dfrac{G_{N,M,P}(t')}{\vert G_{N,M,P}(t')\vert}\vert,
\label{eq:tracedistance}
\end{equation}
to measure the difference between two different coupling matrices. Under fixed network parameters, the influence caused by changing network structure on synchronization can be measured by using average distance $\bar{D}(N,M,P)=(\int^{t}_{t_0} D(t_0,\tau)d\tau)/(t-t_0)$. Here $\vert A\vert$ is defined as $\sqrt{A^\dagger A}$.

As illustrated in Fig. \ref{fig:fig6}(a), the average distance will decrease with the increasing of the quantum coupling intensity $\mu$ for a SW network with $N=12$ and $P=0.1$. Especially corresponding to $\mu=0.02$, $\bar{D}$ is approximately equal to $0.01$ and it will be less than $0.015$ even $P=0.3$ (see inset in Fig. \ref{fig:fig6}(a)). Such a small difference indicates that quantum coupled nodes can be directly synchronized without any additional control in SW network.

In order to verify the above discussion, we calculate dynamical evolution of a $12$-node SW network (Fig. \ref{fig:fig5}(a)). We find that the network synchronization can be more efficient since each node can continue to synchronize other nodes. As an example, here each major node is
known as a central node to derive the star-type network with $S$ secondary nodes (blue points in Fig .\ref{fig:fig5}(a)). In this case, we can still realize synchronously the synchronization among a large number of nodes even in SW network.

In Fig. \ref{fig:fig6}, we show the evolutions of optical field fidelity and field-oscillator entanglement of some key nodes dissipating into different baths. For each linear motif quantum coupling, the fidelity can stabilize at $100\%$ for a long evolution time. Unlike SF network and Erd\"os-R\'enyi random network \cite{28,ad}, we emphasize that all nodes are directly connected to the network without modifying deliberately any parameters. Therefore the state sharing in this structure is quite convenient. Moreover, as we discussed in point-to-point processing, the differences caused by bath influence can be eliminated by the coupling. It is further verified by the Fig. \ref{fig:fig6}(b) that $\mathcal{F}(\infty)\rightarrow 100\%$ is always satisfied whether in SB, CB or LB. Fig.\ref{fig:fig6}(c) and (d) illustrate that consistent quantum states belonging to different nodes are not thermal equilibrium states with meaningless but have significant quantum properties. The optical field and oscillator in each node are still entangled even $t\rightarrow 7500$, and simultaneously, the negative is consistent between nodes in the same linear motif. 

Now we discuss the synchronization performance when network structure is constantly changing. Our aim is to prove that a state sharing processing will not be affected by other processes in the same network. We consider such a process: two nodes are continuously connected (red dot box in Fig. \ref{fig:fig5}(b)), in the meantime, other links (blue box in Fig. \ref{fig:fig5}(b)) are disconnected at $t_1$, and reconnected at $t_2$ (see dotted boundary lines in Fig. \ref{fig:fig6}(e) and (f)). It can be found from Fig. \ref{fig:fig6}(e) that the nodes with continuous connection will continue to maintain synchronization and $\mathcal{F}(\infty)\rightarrow 100\%$ whether other links are connected or disconnected. It can be predicted that identical quantum states can be transmitted in this SW network even though its topology is varying with time. Similarly, we emphasize again that any parameter correction or additional control on nodes is not needed in this varying network. Therefore, any links can be established or broken off at any time. Moreover, we show that the disconnected nodes can also be synchronized when they are connected together again in Fig. \ref{fig:fig6}(f). The consistent evolution trends corresponding to fidelity  and synchronization measure explain that quantum state sharing is present when synchronization arises. 

\subsection*{Synchronization in scale-free network}
\begin{figure}[]
\centering
\includegraphics[width=6in]{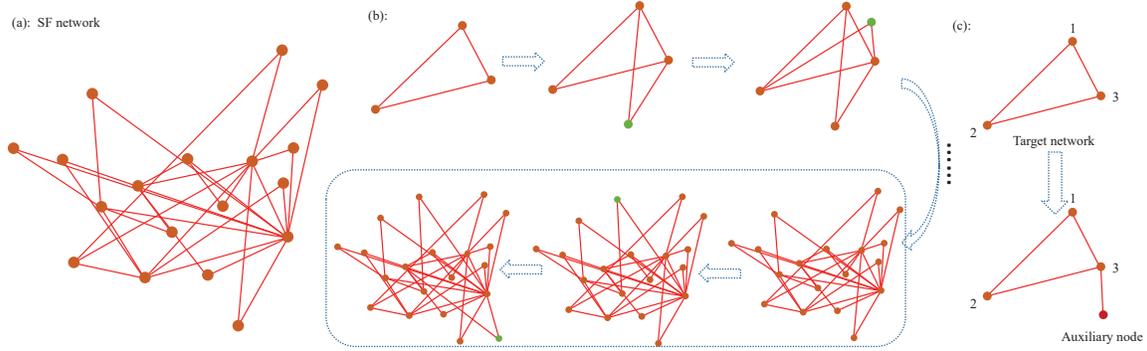}  
\caption{\textbf{Schematic diagram of SF network}. (a): Structure diagram of a SF network with $18$ nodes. (b): A SF network with variable structure. (c): A simple example of network synchronization with auxiliary node. Here the red lines represent the quantum couplings.
\label{fig:fig7}}
\end{figure}
The same protocol discussed above can be applied to synchronize few elements (two or three major nodes) of a network. However, it can not ensure that lots of nodes can be synchronized simultaneously via complex connections. In particular, if connected node constantly connects to other nodes or the probability of edge adding is not small, the whole network will become more complex and lose its symmetry. In this case, the whole network can not be synchronized just through connecting directly nodes each other. The potential scheme to synchronize asymmetric network requires to modify the parameters of each local node. 

To illustrate the problem, we consider a typical SF network as shown in Fig. \ref{fig:fig7}(b). It starts from the network with $m_0$ nodes and in each subsequent process, a new node is added and connected to the network with $m$ existing nodes ($m<m_0$). What need to be explained is that the new node is apt to connect with such nodes with large node degree. Different from the SW network, here time-dependent topology refers to that new nodes access to network constantly.

In order to synchronize all nodes in such a SF network, here we introduce and adopt the dissipative condition, which can be expressed as
\begin{equation}
\sum^N_{k=1}g^q_{jk}=0
\label{eq:dissipativecondition}
\end{equation}
to control network synchronization \cite{55}. Here $g^q_{jk}$ is the element of coupling matrix $G^q$ which describes a special coupling structure. Once the dissipative condition is satisfied, all quantum coupling terms in dynamics equation will no longer exist when all nodes have been synchronized. In order to explain this condition more clearly, let us rewrite Eq. \eqref{eq:dissipativecondition} as $g_{jj}=-\sum^N_{k=1,k\neq j}g_{jk}$ and discuss its both sides, respectively. Considering the quantum Langevin equations, non-diagonal elements of the coupling matrix $G^q$ correspond to the beam splitter (BS) terms $-\mu_{jk}(b^{\dagger}_jb_k+b^{\dagger}_kb_j)$ and reversely, diagonal elements $g_{jj}$ can be regard as frequency shift terms $-\Delta\omega_jb^{\dagger}_jb_j$ in Hamiltonian. Mathematically, those shift terms can express diverse frequencies of different nodes via setting a uniform reference frequency $\omega_{s}$, i.e., $\omega_{mj}=\omega_{s}+\Delta\omega_j$. Besides the dissipative condition, identical node function requires identical $\eta$ for each node. Combining with these two requirements, we gain the conditions for whole network synchronization as follows:
\begin{equation}
\Delta\omega_j=\sum^N_{k=1}\mu_{jk}\,\,\,\,\,\,\,\,\,\,\,\,\,\,\,\,\,\,\,\,\,\,\&\,\,\,\,\,\,\,\,\,\,\,\,\,\,\,\,\,\,\,\,\,\,\dfrac{Q_{MR,j}}{\omega_s-\Delta\omega_j}=constant,
\label{eq:finnalcondition}
\end{equation}
where $\omega_s$ should meet $\omega_s <\min\{\omega_{mj}\}$ owing to $\mu_{jk}>0$. In this case Eq. \eqref{eq:Daanetwork} will reduce to 
\begin{equation}
\begin{split}
&\dot{A}_i=[-\kappa+i\Delta_i+ig(B^{*}_i+B_i)]A_i+E\\ 
&\dot{B}_i=\left\{-\gamma-i\omega_{s}[1+\dfrac{C_i(t)}{2}]\right\}B_i+ig\vert A_{i}\vert^2-i\dfrac{\omega_s}{2}C_i(t)B^{*}_i,
\label{eq:Deffnetwork}
\end{split}
\end{equation}
owing to $B_i=B_j$ ($j\in[1,N], j\neq i$). Eq. \eqref{eq:Deffnetwork} implies that synchronization state can be determined only by respective node functions which have the same forms due to synchronized voltage control. The network evolution under this mechanism can be interpreted as follows: Different initial states and node functions will lead to initial differences among all node, which can be regarded as phase differences when the node functions are controlled with identical parameters. In phase space, mutual quantum coupling makes the error between two nodes take on periodic evolution in weak nonlinear regime, which ensures that the systemic evolutionary track is a limit cycle but not chaos. Therefore, there is always such a moment (the least common multiple of the error periods) that all node errors simultaneously tend to zero. Once this phenomenon emerge, zero coupling terms and identical node functions ensure the system variables to sustain synchronous state.

Note that the first equation in Eq. \eqref{eq:finnalcondition} is non-idempotent linear equation and it can not always be solved since Hermitian BS terms require $\mu_{jk}=\mu_{kj}$. To ensure there exists at least one solution, a simple method is to add some auxiliary nodes and connections into the network which should be designed for satisfying the solution condition. In other words, one can always find suitable parameters to ensure state sharing within the whole complex network.

In order to explain the synchronization conditions more intuitively, we discuss a triangle structure as an example to show how to adjust the coupling strength to satisfy the synchronization conditions (see Fig. \ref{fig:fig7}(c)). Considering this kind of structure, Eq. \eqref{eq:finnalcondition} becomes
\begin{equation}
\begin{split}
&\omega_{m1}-\omega_s=\mu_{12}+\mu_{13}\\ 
&\omega_{m2}-\omega_s=\mu_{12}+\mu_{23}\\ 
&\omega_{m3}-\omega_s=\mu_{23}+\mu_{13}.
\label{eq:examplene}
\end{split}
\end{equation}
In above expressions, $\omega_{m1,2,3}$ are oscillator frequencies of target nodes and they should be arbitrary but fixed. Relatively, $\omega_s$ is a designed standard frequency and it is similar to $\mu_{12,13,23}$ which can be adjusted according to different coupling structures or network parameters. Under normal circumstances, there may not exist physical solutions for the equation while $\omega_{m1,2,3}$ are taken arbitrarily. As mentioned above, we add an auxiliary node and connection into the network to ensure the synchronization conditions can be satisfied. Considering the auxiliary node, Eq. \eqref{eq:examplene} becomes
\begin{equation}
\begin{split}
&\omega_{m1}-\omega_s=\mu_{12}+\mu_{13}\\ 
&\omega_{m2}-\omega_s=\mu_{12}+\mu_{23}\\ 
&\omega_{m3}-\omega_s=\mu_{23}+\mu_{13}+\mu_{A}\\
&\omega_{mA}-\omega_s=\mu_{A}.
\label{eq:exampleneqq}
\end{split}
\end{equation}  
where $\omega_{mA}$ is the oscillator frequency of auxiliary node. Eq. \eqref{eq:exampleneqq} will exist infinitely many solutions by properly selecting $\omega_A$ and $\mu_A$. In particular, if we set
\begin{equation}
\begin{split}
&\omega_s=\omega_{m1}-\mu_{12}-\mu_{13}\\
&\mu_{23}=\mu_{13}+\omega_{m2}-\omega_{m1}\\
&\mu_{A}=\mu_{12}-\mu_{13}+\omega_{m3}-\omega_{m2}\\
&\omega_{mA}=-2\mu_{13}+\omega_{m1}-\omega_{m2}+\omega_{m3},
\label{eq:exampleneqq}
\end{split}
\end{equation} 
the dissipative condition will be satisfied even for arbitrary $\mu_{12}$ and $\mu_{13}$.

In order to validate that above ideas can be extended into complex networks, here we calculate dynamical evolution of a $18$-node SF network (Fig. \ref{fig:fig7}(c)). Quantum sharing within the network is measured and analyzed by a network-averaged fidelity which is defined as 
\begin{equation}
\mathcal{F}(N)=\dfrac{\sum_{j<k}\mathcal{F}_{jk}}{C_N^2}.
\label{eq:networkaveragedfidelity}
\end{equation}
Here $\mathcal{F}_{jk}$ is the fidelity of nodes $j$ and $k$ and $C^2_N$ is the combination number. Under this definition, that $\mathcal{F}(N)$ tends to $100\%$ means that the fidelity between any two nodes tends to a hundred percent.

\begin{figure}[]
\centering
\includegraphics[width=6in]{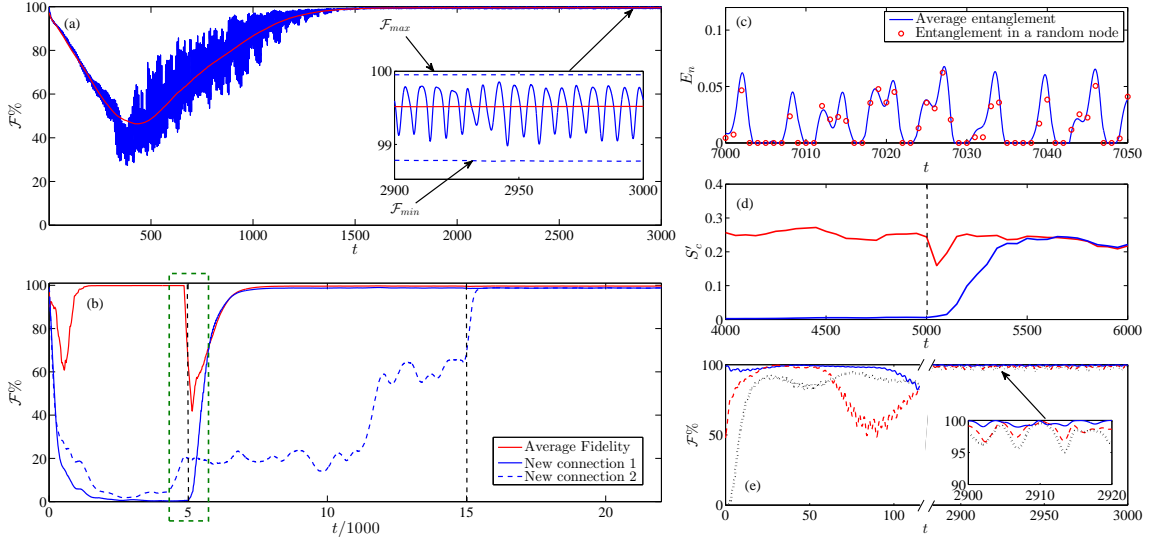}  
\caption{\textbf{Evolutions of fidelity and entanglement in SF network.} (a): Network-averaged fidelity (blue) and its local time average value (red). (b): Fidelities between two nodes when the network structure varies. (c): Entanglement measures of the quantum states in each cavity. (d): Quantum synchronous measures when the network structure varies and the time interval corresponds to the green box in (b). In (a,b,c,d), the initial state of each cavity field is vacuum state and other initial conditions are all random. (e): Initial state dependency of fidelity. Here the initial states of cavity fields are respectively selected as random coherent states. All the other parameters are the same with Fig. \ref{fig:fig2}.
\label{fig:fig8}}
\end{figure}
In Fig. \ref{fig:fig8}(a), we show the evolution of network-averaged fidelity. It can be seen that $\mathcal{F}(N)>99\%$ is always tenable and time-averaged fidelity can achieve ${\mathcal{F}}=99.53\%$ under the same parameters. Further more, the best result of quantum state sharing between two nodes can achieve ${\mathcal{F}}=99.99\%$, and correspondingly, the worst synchronization effect between two nodes in this network can also ensure a successful state sharing with ${\mathcal{F}}>98.5\%$. 

Now we focus on varying network structure. We consider joining two new nodes into the network one after another (corresponding to blue dashed box in Fig. \ref{fig:fig7}(b)). When new nodes are added, the parameters of network will be adjusted to satisfy synchronization conditions under new structure. In general case, only the nodes which are directly connected with the new joining node need to be adjusted since the SF network requires new nodes to be added on the basis of original structure. In Fig. \ref{fig:fig8}(b), we exhibit that new nodes are not synchronized with other nodes when they are in the free evolution, and the fidelities between new node and other nodes in the network remain at a lower value. Once a node accesses into network (i.e., $t/1000=5$ in Fig. \ref{fig:fig8}(b)), it will synchronize with other nodes rapidly and the corresponding fidelity will continue to rise until almost hundred percent. For other nodes in the network, average fidelity shows that new node will not affect the original synchronization effect although average fidelity decreases when new nodes just access into network. We can also get the same conclusion by studying the synchronization measure of oscillators. As shown in Fig. \ref{fig:fig8}(d), $S_c'$ between new node and any other nodes has a significant improvement and finally stabilized at $0.25$, roughly. To summarize, we can ensure that the state sharing processing can also work even if the network structure is varying. Finally, we plot the entanglement measures to illustrate the nonclassical effect of the field in each node in Fig. \ref{fig:fig8}(c) and show that $\mathcal{F}(\infty)\rightarrow 100\%$ can always be achieved even starting evolution from arbitrary initial state in Fig. \ref{fig:fig8}(e).

\section*{Discussion}

Here we give a brief discussion about feasibility of quantum synchronization in complex network, including the experimental feasibility of coupled optomechanical array and the solvability of network synchronization conditions corresponding to Eq. \eqref{eq:finnalcondition}. In our simulations, we select the same dimensionless parameters adopted in Mari's work and they can meet the experimental requirements of most optomechanical systems. Especially in membrane (nano-object)
 optomechanical system, the parameter bounds are: $\omega_m/(2\pi)=134$ kHz, $m_{eff}=40\sim150$ng, $\gamma=4$ kHz and $\kappa=157$kHz \cite{56,57,58,59}. Correspondingly, $g\sim\gamma<\kappa$ is also a common coupling intensity in recent researches of optomechanical system \cite{5,13,14,41,42}. Besides, phonon coupling between charged oscillators is also widely discussed in oscillator synchronization and optomechanical array \cite{5,14}. 
 
In summary, we have proposed and analyzed a quantum state sharing process in complex quantum network. The quantum channel of this state sharing process is based on quantum synchronized oscillators and it will keep strong quantum correlation in Markovian dissipation for a long time. In particular, we have discussed two typical complex networks to illustrate the effectiveness of our state sharing process. SW network can be regarded as independent communication between nodes in a network and the state sharing occurs in linear motifs. Correspondingly, SF network describes an information diffusion process and all nodes need to be synchronized simultaneously in the network. By analyzing quantum synchronization, we find that linked nodes can directly achieve synchronization in small-world network, but the whole network can be synchronized only if the nodes are locally regulated to satisfy given synchronization conditions. In this case, $\mathcal{F}\rightarrow 100\%$ can be achieved in both SW and SF networks and basically it is not influenced by the environment. For SW network, a node can connect (disconnect) other nodes at any time without affecting the other links; For SF network, external nodes can join the network at any time to obtain the same quantum state.  These two properties ensure that our scheme is effective for QIP in complex network. Furthermore, we have also given a brief discussion about the experimental feasibility of our scheme. 

We think some open aspects are worth being further investigated. For examlpes, unknown synchronous quantum state can be used to encrypt quantum states in the quantum state transfer process. Moreover, transmission of continuous variable signal in synchronized optomechanical system was discussed in Ref. \citen{14}. Because the synchronization and state sharing are both effective in SW and SF netwroks, we think the scheme about quantum synchronization and state sharing discussed in our work can exhibit potential application values in communication network.

\section*{Appendix}

\subsubsection*{Analysis of point-to-point electro-optomechanical system}
In this appendix, we will give the details of the calculation method for obtaining the evolutions of covariance matrix, Gaussian negativity and Gaussian fidelity. After transforming the annihilation and creation operators by using $\hat{x}_j=(a^{\dagger}_j+a_j)/\sqrt{2}$, $\hat{y}_j=i(a^{\dagger}_j-a_j)/\sqrt{2}$, $\hat{q}_j=(b^{\dagger}_j+b_j)/\sqrt{2}$ and $\hat{p}_j=i(b^{\dagger}_j-b_j)/\sqrt{2}$, Eq. (\ref{eq:fluctuations}) can be rewritten in a more compact form: $\partial_t\hat{u}=S\hat{u}+\hat{\xi}$. Here vector $\hat{u}$ is defined as $\hat{u}=(\delta x_1, \delta y_1, \delta x_2, \delta y_2, \delta q_1, \delta p_1, \delta q_2, \delta p_2)^{\top}$ and $\hat{\xi}$ means input vector $(\hat{x}^{in}_1, \hat{y}^{in}_1, \hat{x}^{in}_2, \hat{y}^{in}_2, \hat{q}^{in}_1, \hat{p}^{in}_1, \hat{q}^{in}_2, \hat{p}^{in}_2)^{\top}$. $S$ is a time-dependent coefficient matrix 
\begin{equation}
S=
\begin{pmatrix}
 -\kappa & -\Gamma_1& 0&  0&  -2g\text{Im}[A_1]& 0&  0&0 \\ 
\Gamma_1&  -\kappa&  0&  0&  2g\text{Re}[A_1]&  0&  0&0 \\ 
 0&  0& -\kappa & -\Gamma_2 &0&  0&  -2g\text{Im}[A_2]&  0 \\ 
 0&  0&  \Gamma_2&  -\kappa&  0&  0&  2g\text{Re}[A_2]&0 \\ 
 0&  0&  0&  0&  -\gamma&  \omega_{m1}&  0&-\mu \\ 
 2g\text{Re}[A_1]&  2g\text{Im}[A_1] &  0&  0&  -\omega_{m1}[1+C_j(t)]&  -\gamma&  \mu& 0\\ 
 0&  0&  0&  0&  0&  -\mu&  -\gamma& \omega_{m2}\\ 
 0&  0&  2g\text{Re}[A_2]&  2g\text{Im}[A_2]&  \mu&  0& -\omega_{m2}[1+C_j(t)]& -\gamma
\end{pmatrix}
\end{equation}
which contains the information of the mean value. Where $\Gamma_j=\Delta_j+2g\text{Re}[B_j]$. In order to analyze quantum synchronization and quantum correlation, we consider following covariance matrix 
\begin{equation}
V_{ij}(t)=V_{ji}(t)=\dfrac{1}{2}\langle \hat{u}_{i}(t)\hat{u}_{j}(t)+\hat{u}_{j}(t)\hat{u}_{i}(t)\rangle
\label{eq:covariance}
\end{equation}
and its evolution satisfies \cite{35}
\begin{equation}
\partial_tV=SV+VS^{\top}+N.
\label{eq:covarianceevolution}
\end{equation}
Here $N$ is a diagonal noise correlation matrix $N_{ij}\delta(t-t')=\langle\hat{\xi}_{i}(t)\hat{\xi}_{j}(t')+\hat{\xi}_{j}(t')\hat{\xi}_{i}(t)\rangle/2$. According to Eqs. (\ref{eq:Duffingcoupling}), (\ref{eq:contorlexex}), (\ref{eq:meanvalue}) and (\ref{eq:covarianceevolution}), the first-order synchronization measure between two oscillators can be obtained by $\langle q_-\rangle=\text{Re}[B_1]-\text{Re}[B_2]$ and $\langle p_-\rangle=\text{Im}[B_1]-\text{Im}[B_2]$. Besides, the second-order synchronization measure is:
\begin{equation}
\begin{split}
S'_c(t)=&\langle\delta \hat{q}_{-}^2(t)+\delta \hat{p}_{-}^2(t)\rangle^{-1}\\
=&\left\{\dfrac{1}{2}[V_{55}(t)+V_{77}(t)-2V_{57}(t)]+\dfrac{1}{2}[V_{66}(t)+V_{88}(t)-2V_{68}(t)]\right\}^{-1}.
\label{eq:Scenv}
\end{split}
\end{equation}
The initial state of the optical field is set as a Gaussian vacuum state and the linear interaction will make sure that optical field is always a Gaussian state with the evolution. The fidelity of two general Gaussian states can be obtained using the following formula \cite{51,52}:
\begin{equation}
\begin{split}
\mathcal{F}=\dfrac{2}{\sqrt{\Lambda+\lambda}-\sqrt{\lambda}}\exp[-\beta^{\top}(V_1+V_2)^{-1}\beta],
\label{eq:fidelity}
\end{split}
\end{equation}
where 
\begin{equation}
V1=
\begin{pmatrix}
 V_{11} & V_{12} \\ 
 V_{21} & V_{22}
\end{pmatrix}, 
\,\,\,\,\,\,\,V2=
\begin{pmatrix}
 V_{33} & V_{34} \\ 
 V_{43} & V_{44}
\end{pmatrix}, 
\,\,\,\,\,\,\,\beta={\sqrt{2}}\begin{pmatrix}
 \text{Re}{A_1}-\text{Re}{A_2} \\ 
 \text{Im}{A_1}-\text{Im}{A_2}
\end{pmatrix}, 
\,\,\,\,\,\,\,\Lambda=\det(V_1+V_2),\,\,\,\,\,\,\,\lambda=(\det{V_1}-1)(\det{V_2}-1).
\end{equation}

Finally, we give the detailed calculation method of the Gaussian entanglement measure. In order to calculate \textit{logarithmic negativity} conveniently, we express the covariance matrix $V$ as following compact form:
\begin{equation}
V=
\begin{pmatrix}
 I_{O1} & D_{O1,O2} & D_{O1,M1} &  D_{O1,M2}\\ 
 D_{O2,O1} & I_{O2} & D_{O2,M1} &  D_{O2,M2}\\
 D_{M1,O1} & D_{M1,O2} & I_{M1} &  D_{M1,M2}\\ 
 D_{M2,O1} & D_{M2,O2} & D_{M2,M1} & I_{M2}
\end{pmatrix}, 
\end{equation}
where $I_i$ and $D_{ij}$ are $2\times2$ matrices. Here we use the indices $O$ and $M$ to specify
the mechanical and optical modes, moreover, $(i,j)$ denotes the entanglement between the modes $i$ and $j$. For example, ``$(O_1,M_1)$'' means the entanglement between the optical mode and the  mechanical mode of the system $1$, correspondingly, ``$(M_1,M_2)$''  is the entanglement between the mechanical modes of the systems $1$ and $2$. The covariance matrix of two entangled modes in this case can be written as:
\begin{equation}
\nu_{ij}=
\begin{pmatrix}
 I_{i} & D_{i,j} \\  
 D_{j,i} & I_{j}
\end{pmatrix}, 
\end{equation}
and the \textit{logarithmic negativity} can be calculated based on 
\begin{equation}
E^{i,j}_N=\max[0, -\ln(2\zeta_{ij})].
\end{equation}
In this expression, $\zeta_{ij}$ is the smallest symplectic eigenvalue of the partially transposed covariance matrix $\tilde{\nu}_{ij}$ which can be obtained from ${\nu}_{ij}$ just by taking $p_j$ in $-p_j$ \cite{61,62}. This symplectic eigenvalue can be obtained by calculating the square roots of the ordinary eigenvalues of $-(\sigma\tilde{\nu}_{ij})^2$, where $\sigma=J\oplus J$ and $J$ is a $2\times2$ matrix with $J_{12}=-J_{21}=1$ and $J_{11}=J_{22}=0$.

\section*{Acknowledgements}
All authors thank Dr. Jiong Cheng, Dr. Wenzhao Zhang and Dr. Yang Zhang for the useful discussion. This research was supported by the National Natural Science Foundation of China (Grant No 11175033, No 11574041, No 11505024 and No 11447135) and the Fundamental Research Funds for the Central Universities (DUT13LK05).

\end{document}